# Cobalt-Catalyzed Chain Transfer Polymerization Enables Soft Methacrylate Nematic Elastomers for Switchable Pressure-Sensitive Adhesion

Noboru Koshimizu[1,2], Mohand O. Saed [1]*

Cavendish Laboratory, University of Cambridge, J.J. Thomson Avenue, Cambridge, CB3 0HE, United Kingdom

Electronic and Imaging Materials Res. Labs, Toray Industries, Inc., 3-2-1 Sonoyama, Otsu, Shiga 530-0842, Japan

E-mail: mos29@cam.ac.uk

**ABSTRACT**

Liquid crystal elastomers (LCEs) exhibit unique viscoelastic behavior arising from their reversible liquid-crystalline ordering, making them attractive candidates for switchable pressure-sensitive adhesives (PSAs). However, methacrylate-based LCEs are inherently highly crosslinked, resulting in elevated glass transition temperatures (Tg) and storage moduli (E′) that limit their adhesive performance. Here, we demonstrate that catalytic chain-transfer polymerization provides an effective strategy for engineering soft methacrylate nematic elastomers through systematic control of network architecture. Incorporation of ppm-level concentrations of bis(boron

difluorodimethylglyoximate)cobalt(II) (CoBF) during photopolymerization reduced the effective crosslink density and increased the molecular weight between crosslinks, resulting in substantial decreases in $Tg$ and $E'$ while preserving nematic order. Dynamic mechanical analysis revealed that increasing CoBF concentration enhanced viscoelastic dissipation and broadened the accessible nematic temperature range. To further optimize the rheological properties for pressure-sensitive adhesion, monofunctional methacrylates and flexible poly(ethylene glycol) dimethacrylate (PEGDMA) were incorporated into the network. The optimized formulation exhibited a $Tg$ near 0 °C, a room-temperature storage modulus of approximately 0.3 MPa, and high damping behavior, approaching the Dahlquist criterion for pressure-sensitive adhesion. Consequently, the resulting nematic elastomers exhibited strong tack, peel, and lap-shear adhesion in the nematic state together with rapid, reversible, and residue-free debonding upon heating above the nematic-to-isotropic transition temperature. These results establish catalytic chain transfer as a versatile platform for tuning the structure–property relationships of methacrylate nematic elastomers and provide a scalable route toward reusable and debond-on-demand adhesive materials.

## INTRODUCTION

Liquid crystal elastomers (LCEs) can dynamically alter their physical properties through liquid crystal phase transitions, enabling unique potential applications in reversible actuation, dynamic adhesion, and vibration damping.[1,2] These behaviors originate from the liquid crystalline phase structure: in the nematic phase, mesogens align along a director, and reorientation of this director under deformation introduces internal friction and relaxation processes.[3] This results in a high loss factor ($\tan \delta$, indicating energy dissipation) and a low storage modulus ($E'$), allowing the material to maintain its deformed shape.[4,5] In contrast, in the isotropic phase, the absence of a director eliminates these relaxation contributions, reducing energy dissipation and lowering $\tan \delta$, thereby enabling deformation into an energetically stable shape.[3,4] Consequently, LCEs serve as smart polymers for applications such as damping,[5–7] dynamic adhesives,[8–10] and actuators.[11–15] Pressure-sensitive adhesives (PSAs) are a class of soft and solid polymeric materials that must remain permanently and aggressively tacky at room temperature, adhering to surfaces regardless of their surface energy upon mere contact without the need for more than finger pressure.[16–18] The performance of PSAs depends heavily on their intrinsic bulk properties, particularly their capacity to dissipate energy via viscoelastic mechanisms.[19] Because LCEs exhibit exceptionally high energy dissipation among solid polymers, they are promising candidates for high-performance PSA design;[1,5,7] moreover, the dynamic nature of this dissipation allows these systems to achieve on-demand switchable adhesion (high adhesion in nematic phase and low adhesion in isotropic phase).[8–10,20–23] Conventionally, switchable LCE-based adhesives require pre-application annealing to achieve sufficient surface wetting, a limitation imposed by their high room-temperature shear modulus.[8,24] To overcome this limitation, LCEs must satisfy the Dahlquist criterion. This rule dictates that an adhesive must remain permanently tacky at its operating temperature, ideally exhibiting a glass transition temperature ( $Tg$ ) below 0 °C and a shear modulus ($G'$) less than 0.1 MPa at a frequency of 1 Hz.[16–18,25] We recently achieved these criteria by designing soft side-chain LCEs; however, those systems rely on custom-made mesogens, which limits their scalability.[26]

(Meth)acrylate chemistries dominate the commercial PSAs market, comprising approximately 60–70% of total PSA formulations.[27–29] Their popularity stems from low-cost raw materials, excellent adhesive performance, and straightforward processing via coating and rapid UV curing.[28] Additionally, the vast library of commercial (meth)acrylic monomers allows precise tuning of mechanical, optical, and thermal properties. These low-viscosity liquid monomers blend seamlessly with initiators and crosslinkers,[30,31] undergoing efficient thermal or photoinduced free-radical polymerization.[28] However, the design of (meth)acrylic-based LCE PSAs presents a significant challenge. Radical polymerization of (meth)acrylate reactive mesogens typically produces glassy liquid-crystalline polymer networks with high $Tg \approx 50$ to 200 °C and high $G'$ at room temperature (10 to 1000 MPa).[32,33] These elevated $Tg$ and $G'$ values stem from the intrinsic rigidity of the mesogenic units combined with the high crosslinking density of the network, placing these materials far above the ideal thresholds defined by the Dahlquist criterion. Consequently, satisfying this criterion requires a significant reduction in the crosslinking density of liquid-crystalline polymer networks. One of the most effective strategies for reducing crosslink density in acrylate-based systems is the incorporation chain transfer agents (CTAs). Common examples include mercaptans (thiols), reversible addition–fragmentation chain transfer (RAFT) agents for control radical polymerization, cobalt complexes, and halogenated compounds[34–37]. Mechanistically, CTAs terminate propagating polymer radicals and generate new radical species. This process increases the average molecular weight between crosslinks (Mc), decreases the effective crosslink density, and increase the concentration of dangling ends. These structural modifications enhance chain mobility and free volume while promoting entanglement, ultimately yielding a lower $Tg$, reduced $G'$, and enhanced viscoelastic dissipation.[38–40] For instance, White and co-workers effectively utilized thiol-based CTAs to reduce crosslink density in acrylic liquid-crystalline polymer networks, enabling the fabrication of soft LCE actuators.[41,42] Building on this approach, our group further optimized the thiol-based CTA strategies to develop acrylate-based LCE PSAs that satisfy the Dahlquist criterion. Specifically, incorporation of a non-equimolar thiol CTA loading (2:1 molar ratio of thiol to diacrylate mesogenic monomers) reduced $Tg$ from >200 °C to -10 °C and decreased the room-temperature $E'$ from 4000 to 0.3 MPa, compared to the control formulation without any thiol CTA.[43] However, reducing crosslink density to achieve *Tg* and *E′* values that satisfy the Dahlquist criterion typically requires exceptionally high thiol CTA loadings, often exceeding the concentration of the acrylate monomers themselves. This results in LCE networks with a high concentration of unreacted thiol groups. This limitation arises because only ~50 mol% of the thiol functional groups are incorporated into the final polymer network.[41–44] This low thiol conversion is attributed to competitive kinetics between the thiol–acrylate chain transfer reaction and acrylate homo-polymerization, with the latter strongly favored under typical radical polymerization conditions.[41–45] Therefore, a CTA that can be used in small quantities without leaving unreacted groups is highly desirable. Recently, Worrell et al. employed cobalt(II)-based chain transfer agents (CTAs), such as bis(boron difluorodimethylglyoximate) cobaltate(II) (CoBF), to reduce the crosslinking density in methacrylate-based photopolymer networks.[46] Unlike thiols, cobalt(II) complexes act as highly potent CTAs, enabling a reduction in loading to approximately one ten-thousandth of conventional amounts (i.e. ppm levels).The proposed mechanism, known as catalytic chain transfer polymerization (CCTP), involves abstraction of a β-methyl hydrogen from the growing polymer chain by the Co(II) complex, yielding an alkene-terminated polymer chain and a Co(III)–hydride species. The Co(III)-hydride then transfers a hydrogen atom to a methacrylate monomer, initiating a new chain while regenerating the Co(II) catalyst. This catalytic cycle enables efficient control of network architecture at ppm-level concentrations. Reported results show that *Tg* decreased from 34 °C to −27 °C, *E′* declined from 62.3 MPa to 6.3 MPa, and *tan δ* increased from 0.4 to 0.9.[46] It is important to note that CCTP is effective in methacrylate-based systems but does not operate in acrylate-based materials.[46] Consequently, this strategy is not applicable to conventional acrylate-based LC materials. Inspired by this work, we employed CoBF as a catalytic CTA to reduce the crosslinking density of methacrylate-based LCE photopolymer networks, with the goal of transforming inherently glassy liquid-crystalline polymers into soft PSA materials. Multifunctional methacrylate LC monomers were first combined with CoBF and a photoinitiator and subsequently polymerized under UV irradiation. This process requires no purification and is readily scalable. The LCE adhesive formulations were optimized to satisfy the Dahlquist criterion by adjusting CoBF loading and incorporating monofunctional LC and isotropic methacrylate monomers, while maximizing internal viscoelastic dissipation to enhance adhesive strength. As a result, the optimized LCE PSA exhibited a $T_g$ of –10 °C, an $E'$ of 0.3 MPa at room temperature, and an elevated tan δ within the nematic phase. Crucially, the material demonstrated switchable adhesion, showing strong adhesion at room temperature and nearly zero adhesion at elevated temperatures driven by the liquid-crystal phase transition. To our knowledge, catalytic chain-transfer polymerization has not previously been used to engineer soft methacrylate nematic elastomers for pressure-sensitive adhesion.

## EXPERIMENTAL SECTION

### Materials:

Cobalt(II) acetate tetrahydrate was purchased from Strem Chemicals. Dimethylglyoxime, 4-hydroxybenzoic acid, 6-bromo-1-hexanol, 4-cyano-4′-hydroxybiphenyl, 1-ethyl-3-(3-dimethylaminopropyl)carbodiimide (EDC), and 4-dimethylaminopyridine (DMAP) were obtained from Tokyo Chemical Industry (TCI, Japan). Boron trifluoride diethyl etherate, 1,2-ethanedithiol (EDT), 2,2′-(ethylenedioxy)diethanethiol (EDDET), 2,6-di-tert-butyl-4-methylphenol (BHT), dipropylamine (DPA), triethylamine (TEA), 2-isocyanatoethyl methacrylate (IEM), poly(ethylene glycol) dimethacrylate (PEGDMA, Mn ≈ 750 g mol$^{-1}$), methyl methacrylate (MMA), sodium hydroxide, potassium iodide, hydrochloric acid (37 wt %), methacrylic acid, hydroquinone, p-toluenesulfonic acid monohydrate, sodium sulfate, silica gel (230–450 mesh), potassium bicarbonate, sodium chloride, and 2,2-dimethoxy-2-phenylacetophenone (Irgacure 651) were purchased from Sigma-Aldrich and used as received unless otherwise stated. RM257 was purchased from Chemfish Tokyo Co., Ltd. Diethyl ether ($Et_2O$), methanol, ethanol, anhydrous tetrahydrofuran (THF), chloroform, dichloromethane (DCM), hexane, ethyl acetate, toluene, chloroform-d ($CDCl_3$), and dimethyl sulfoxide-$d_6$ (DMSO-$d_6$) were purchased from Sigma-Aldrich and used as received. A 1 M hydrochloric acid solution was prepared by dilution of concentrated hydrochloric acid (37 wt %) with deionized water.

**Preparation of LCE Formulations**:

Irgacure 651 (0.5 wt %) and bis(boron difluorodimethylglyoximate)cobalt(II) (CoBF) were dissolved in tetrahydrofuran (THF) together with the corresponding dimethacrylate liquid-crystalline oligomer (RM257–EDT–IEM or RM257–EDDET–IEM) under vigorous stirring until a homogeneous solution was obtained. The resulting precursor mixtures were cast into glass-slide cells separated by 1 mm spacers to prepare free-standing films, poured into dog-bone molds for tensile testing, or coated onto oxygen-plasma-treated poly(ethylene terephthalate) (PET) films to fabricate adhesive tapes for peel testing.

For free-standing films, the precursor solution was injected into glass-slide sandwich cells with a thickness of 1 mm. For tensile testing, the formulations were cast into dog-bone molds with a gauge-section width of 2.1 mm, gauge length of 13.8 mm, grip separation of 24.0 mm, and thickness of 0.4 mm. For adhesive tape fabrication, the formulations were screen-printed onto plasma-treated PET substrates using a 0.3 mm coating gap, yielding an adhesive-layer thickness of approximately 93 μm after solvent removal.

All samples were photopolymerized under UV irradiation (365 nm) for 30 min at room temperature and subsequently dried at 85 °C overnight to remove residual solvent. Samples prepared from RM257–EDT–IEM and RM257–EDDET–IEM were designated EDT–n and EDDET–n, respectively, where n denotes the concentration of CoBF (ppm) relative to the mass of liquid-crystalline oligomer.

**Fourier Transform Infrared Spectroscopy (FTIR):**

Fourier transform infrared (FTIR) spectra were collected using a Nicolet iS10 FTIR spectrometer (Thermo Fisher Scientific) equipped with a Smart iTX attenuated total reflectance (ATR) accessory fitted with a diamond crystal. Free-standing samples were analyzed directly without further preparation. Spectra were recorded over the range of 4000–400 $cm^{-1}$ and processed using OMNIC software (Thermo Fisher Scientific).

For photocured samples, the extent of methacrylate conversion was determined from the decrease in the characteristic C=C stretching vibration relative to the corresponding uncured formulation. Reported conversion values represent the average of at least three independent measurements.

**Gel Fraction:**

The gel fraction was determined by solvent extraction in toluene. Prior to extraction, samples were dried under vacuum and weighed to obtain the initial mass (Wi). Samples were then immersed in toluene at 25 °C for 48 h to extract soluble species. Following extraction, the samples were removed from the solvent, dried at 130 °C for 8 h to constant mass, and reweighed to obtain the final mass (Wf). All measurements were performed in triplicate, and the average value is reported. The gel fraction (GF) was calculated according to:

$$\mathrm{GF}\ (\%) = \frac{W_f}{W_i}\ x\ 100$$

Where $W_i$ is initial dry mass before solvent extraction and $W_f$ is the dry mass after solvent extraction.

**Dynamic scanning calorimetry (DSC):**

Thermal transitions were characterized using a DSC 4000 differential scanning calorimeter (PerkinElmer). Samples (approximately 10 mg) were sealed in standard aluminum pans and subjected to three consecutive heating–cooling cycles under a nitrogen atmosphere. Samples were heated from room temperature to 150 °C at a rate of 10 °C $min^{-1}$, held isothermally for 5 min, cooled to −50 °C at 10 °C $min^{-1}$, and held for an additional 5 min before reheating. Thermal transition temperatures were determined from the third heating cycle to eliminate thermal history effects.

The glass transition temperature (*Tg*) was determined from the midpoint of the heat-capacity step change, while the nematic-to-isotropic transition temperature (*Tni*) was determined from the minimum of the corresponding endothermic peak.

**Dynamic mechanical analysis (DMA):**

Dynamic mechanical analysis (DMA) was performed using a Discovery DMA 850 instrument (TA Instruments). Free-standing samples with a thickness of approximately 1.0 mm were analyzed in tensile mode at a fixed frequency of 1 Hz and a strain amplitude of 0.02%, within the linear viscoelastic regime. Temperature sweeps were conducted from −50 to 150 °C at a heating rate of 2 °C $min^{-1}$.

The storage modulus ($E'$), loss modulus ($E''$), and loss factor ($tan\ \delta$) were recorded as a function of temperature. *Tg* was determined from the peak in the $tan\ \delta$ curve, while the *Tni* was identified from the corresponding change in viscoelastic response associated with the loss of liquid-crystalline order.

**Tensile Mechanical Testing:**

Tensile mechanical properties were characterized using either a Tinius Olsen ST1 universal testing machine or a Discovery DMA 850 (TA Instruments) equipped with a tensile fixture.

For monotonic tensile testing, dog-bone specimens (gauge length: 13.8 mm, gauge width: 2.1 mm, grip separation: 24.0 mm, thickness: 0.4 mm) were stretched at a crosshead speed of 12 mm $min^{-1}$ at 20 °C until failure. Engineering stress–strain curves were used to determine the Young's modulus, yield stress, tensile strength, and elongation at break.

Elastic recovery was evaluated using cyclic tensile testing. Free-standing specimens (width: 3.0 mm, thickness: 1.0 mm) were subjected to a loading–unloading cycle from 0 to 100% strain and subsequently returned to 0% strain at a strain rate of 60% $min^{-1}$ at 22 °C. The loading and unloading curves were used to assess elastic recovery, hysteresis, and residual deformation.

**Probe Tack Test:**

Probe tack measurements were performed using a Tinius Olsen universal testing machine equipped with a spherical stainless-steel probe (diameter = 10 mm). Prior to testing, the probe was cleaned with acetone and brought into contact with the surface of a free-standing LCE film (thickness ≈ 1 mm) supported on a glass substrate. The sample temperature was controlled using a temperature-regulated hot plate. The probe was lowered onto the sample until a compressive force of 0.5 N was reached and maintained for a dwell time of 300 s to ensure intimate contact between the probe and adhesive surface. The probe was subsequently withdrawn at a constant rate of 10 mm $min^{-1}$ while the normal force was continuously recorded. The maximum tensile force observed during debonding was defined as the tack force ($F_{ad}$).

At least three independent measurements were performed for each condition, and the reported values represent the average ± standard deviation.

**180-degree Peel Test:**

Adhesive performance under peeling deformation was evaluated using a 180° peel test performed on a Discovery DMA 850 instrument (TA Instruments). Adhesive tapes were prepared by coating the LCE formulation onto poly(ethylene terephthalate) (PET) film, yielding an adhesive-layer thickness of approximately 93 μm and a tape width of 5 mm. The adhesive tape was applied to either glass or stainless-steel substrates under manual pressure at room temperature to form a three-layer laminate consisting of the substrate, adhesive layer, and PET backing. Prior to testing, samples were equilibrated at the desired temperature for 5 min. The free end of the PET backing was folded back by 180° and clamped in the upper grip, while the substrate was secured in the lower grip. Peel tests were performed at a constant crosshead speed of 100 mm $min^{-1}$, and the peel force was continuously recorded as a function of displacement.

The reported peel strength was calculated by normalizing the measured force by the tape width (N $m^{-1}$). At least three independent measurements were performed for each condition, and the reported values represent the average ± standard deviation.

**Lap-shear Test:**

Lap-shear adhesion measurements were performed using a Tinius Olsen ST1 universal testing machine. Free-standing LCE samples were cut into square specimens (10 × 10 $mm^2$) and placed between two stainless-steel substrates to form a single-lap joint. The assembly was compressed under manual pressure at room temperature to establish adhesive contact.

The bonded substrates were mounted in the testing frame and subjected to uniaxial shear deformation at a constant crosshead speed of 1.25 mm $min^{-1}$ until failure. Force and displacement were continuously recorded during testing. The lap-shear strength was calculated by dividing the maximum load by the bonded area.

Temperature-dependent lap-shear measurements were performed by heating the bonded assembly during testing. Measurements were conducted in both the nematic and isotropic phases to evaluate the effect of the liquid-crystal phase transition on adhesive performance. At least three independent measurements were performed for each condition, and the reported values represent the average ± standard deviation.

## RESULTS AND DISCUSSION

### Synthesis and Network Formation

The objective of this study was to reduce the effective crosslink density of inherently highly crosslinked methacrylate-based nematic elastomer networks to enable switchable pressure-sensitive adhesion. Because cobalt-mediated catalytic chain-transfer polymerization is ineffective in acrylate-based liquid-crystalline mesogens, methacrylate-functionalized liquid-crystalline oligomers were employed as the polymerizable building blocks.[46]

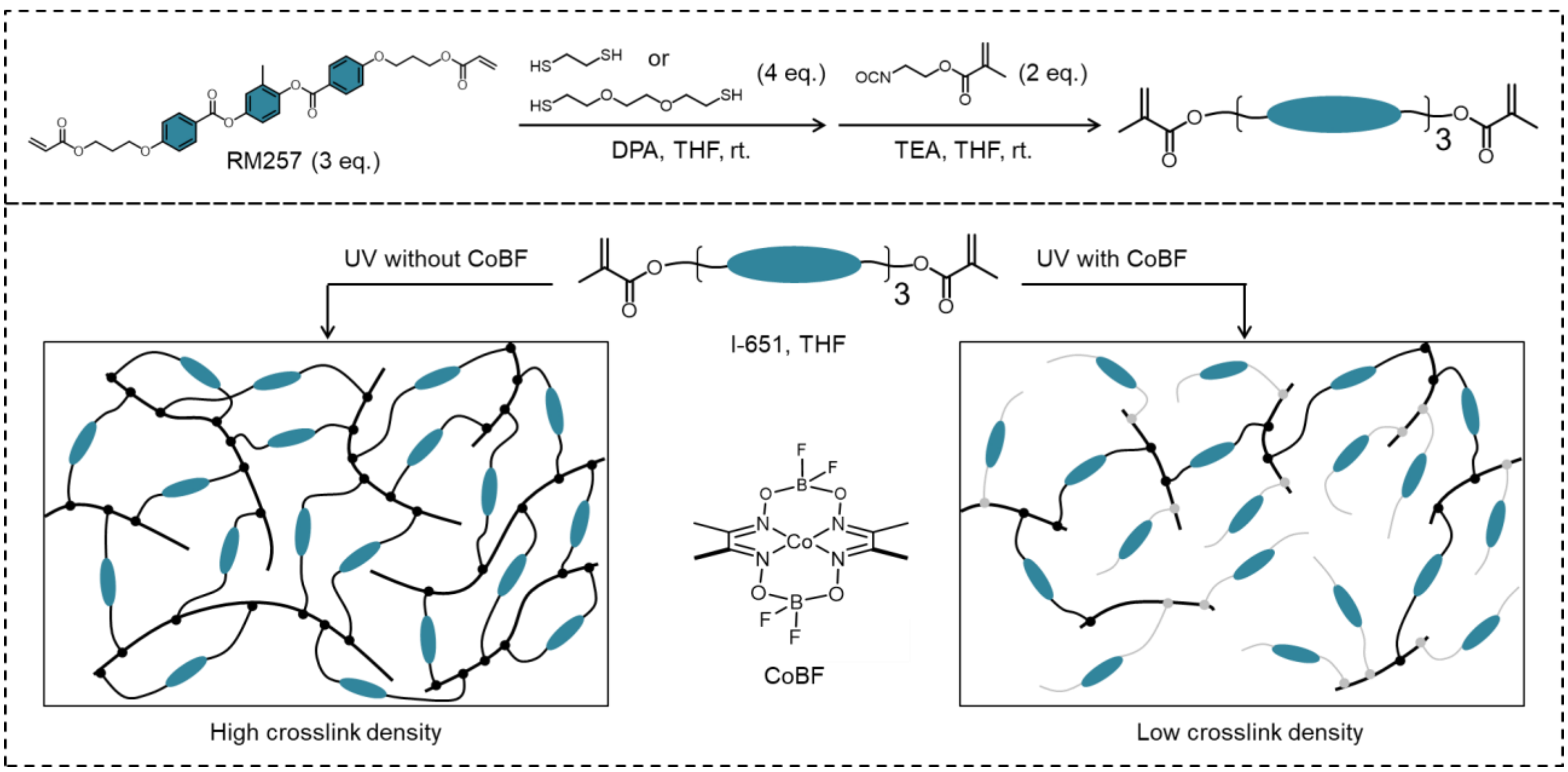

**Scheme 1**. Synthesis of dimethacrylate liquid-crystalline oligomers and schematic illustration of network architectures formed in the absence and presence of catalytic chain-transfer agent (CoBF). Catalytic chain transfer increases the molecular weight between crosslinks and reduces the effective crosslink density, resulting in softer nematic elastomer networks with enhanced chain mobility.

Two dimethacrylate liquid-crystalline oligomers containing identical mesogenic cores but different spacer chemistries were synthesized to investigate the influence of network architecture on thermomechanical properties. Ethanedithiol (EDT) and 2,2′-(ethylenedioxy)diethanethiol (EDDET) were selected as spacers to provide distinct chain lengths and flexibilities. As illustrated in Scheme 1, dithiol-terminated liquid-crystalline oligomers were first prepared through thiol–acrylate Michael addition between RM257 and the corresponding dithiol spacer. Subsequent thiol–isocyanate coupling with 2-isocyanatoethyl methacrylate yielded the dimethacrylate-functionalized oligomers RM257–EDT–IEM and RM257–EDDET–IEM. The chemical structures of all intermediates and final products were confirmed by $^{1}$H NMR and FTIR spectroscopy (Figures S1–S2). To reduce the polymer network crosslinking density, bis(boron difluorodimethylglyoximate)cobalt(II) (CoBF) was employed as a catalytic chain-transfer agent. During methacrylate polymerization, CoBF promotes chain-transfer reactions that terminate growing polymer radicals and generate new initiating species, thereby increasing the molecular weight between crosslinks while decreasing the effective crosslink density.[46] Consequently, CoBF provides a convenient route to systematically tune network architecture using only ppm-level catalyst concentrations. The influence of CoBF loading on network formation was first evaluated using FTIR spectroscopy and gel-fraction analysis. Increasing CoBF concentration from 0 to 175 ppm progressively reduced methacrylate conversion and gel fraction for both EDT- and EDDET-based systems (Tables S1 and S2). This behavior is consistent with increased chain-transfer activity, which suppresses propagation of the chain polymerization and reduces the number of crosslinking points, leading to a higher molecular weight between these points. As expected, the methacrylate conversion decreased with increasing CoBF loading, dropping from 100% at 0 ppm CoBF to 78% at 175 ppm CoBF. This reduction is likely due to the formation of polymer chains terminated with unreacted methacrylate bonds. Interestingly, all formulations exhibited high gel fractions (>85%), confirming successful network formation. At CoBF concentrations above 200 ppm, free-standing films could not be obtained because extensive chain-transfer reactions inhibited network formation and produced insufficiently crosslinked materials. This outcome aligns with the results reported by Bagnall et al.[46] Polarized optical microscopy was subsequently employed to determine whether catalytic chain transfer disrupted liquid-crystalline order. Upon stretching, all samples exhibited maximum birefringence at 45° relative to crossed polarizers and extinction when aligned parallel to the polarizer axis (Figure 1), confirming that nematic ordering was retained despite substantial changes in network architecture.

**Effect of Catalytic Chain Transfer on Thermomechanical Properties:**

To determine how catalytic chain transfer influences the thermomechanical behavior of methacrylate nematic elastomers, the thermal transitions and viscoelastic properties of EDT-n and

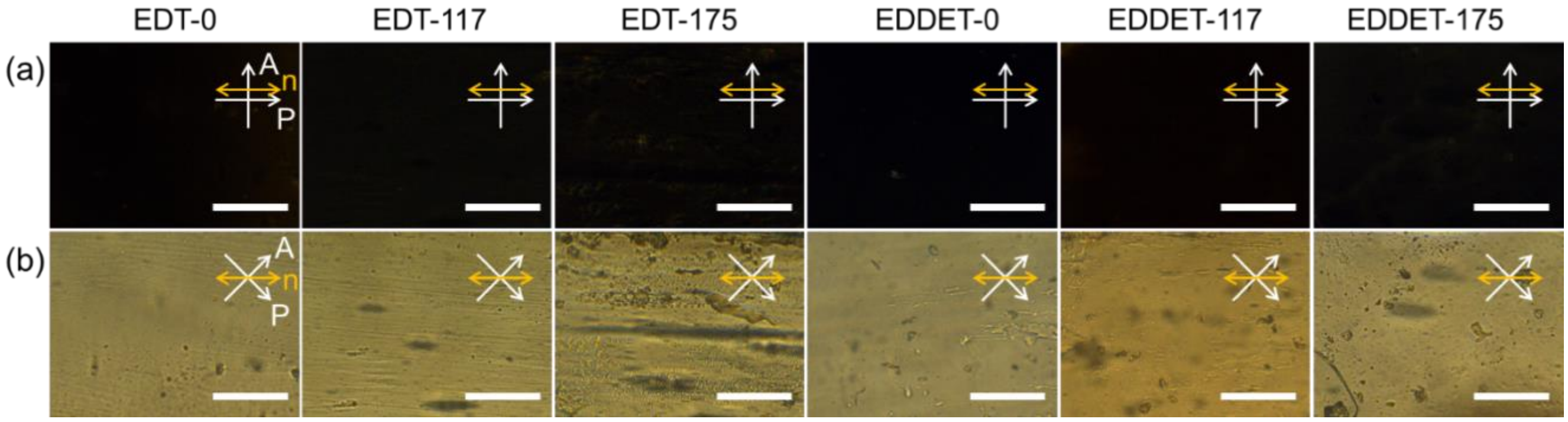

**Figure 1.** Polarized optical microscopy (POM) images of uniaxially aligned methacrylate nematic elastomers containing varying concentrations of CoBF. Images were collected under crossed polarizers with the sample oriented at 45° and 0° relative to the polarizer axis. All samples exhibit birefringence and extinction behavior characteristic of nematic liquid-crystalline order, demonstrating that catalytic chain transfer does not disrupt mesogen alignment or nematic phase formation. (Scale bars: 300 μm).

EDDET-n networks were investigated using DSC and DMA (Figure 2). The incorporation of CoBF produced substantial changes in both the glass transition temperature ($Tg$) and storage modulus ($E'$), indicating significant alterations in network architecture. Representative DSC traces

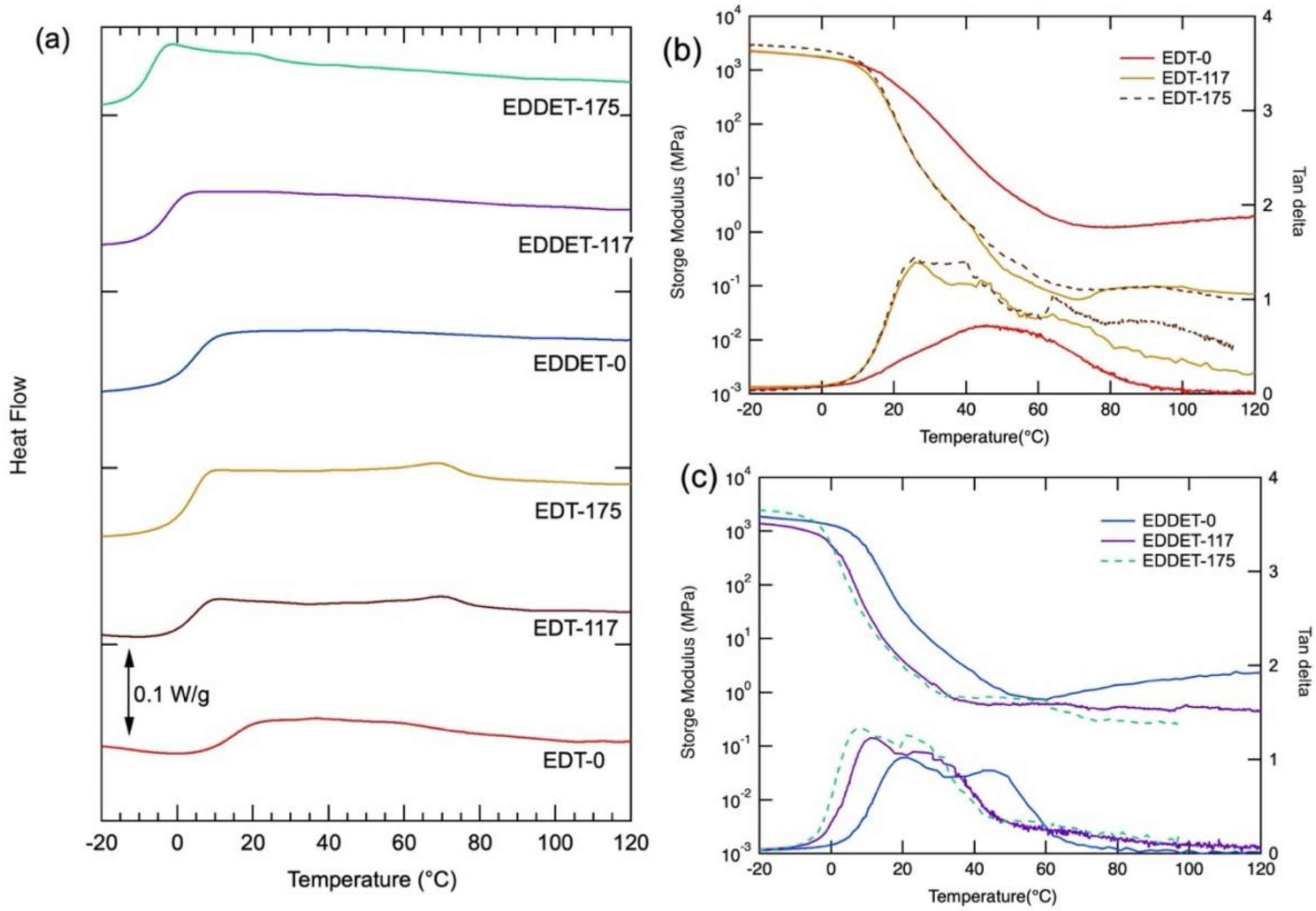

**Figure 2.** Thermomechanical characterization of methacrylate nematic elastomers containing varying concentrations of CoBF. (a) DSC thermograms showing *Tg* and *Tni.* (b) Storage modulus ($E'$) as a function of temperature. (c) Loss factor (*tan δ*) as a function of temperature

of EDT-n samples are shown in Figure 2a. Increasing CoBF loading reduced *Tg* from 14 °C for EDT-0 to 4 °C for EDT-175, while the nematic-to-isotropic transition temperature (*Tni*) remained nearly unchanged at approximately 70 °C. As a result, the temperature range over which the materials remained in the nematic phase increased substantially with increasing CoBF concentration. The broad and poorly defined nematic transition observed for EDT-0 is attributed to the restricted chain mobility associated with its highly crosslinked network structure. A similar trend was observed for EDDET-n samples. Owing to the longer and more flexible EDDET spacer, all EDDET-based networks exhibited lower *Tg* and *Tni* values than the corresponding EDT-based materials. In particular, EDDET-175 exhibited the lowest *Tg* (−8 °C), demonstrating that both spacer flexibility and catalytic chain transfer contribute to enhanced chain mobility. The viscoelastic response of the networks was further examined using DMA (Figure 2b, c). Increasing CoBF concentration systematically shifted the *tan δ* peak to lower temperatures and reduced the storage modulus throughout the measured temperature range. These changes are consistent with catalytic chain-transfer reactions increasing the molecular weight between crosslinks and generating additional dangling-chain segments, thereby reducing the effective crosslink density. The resulting increase in free volume and chain mobility lowers both *Tg* and *E′* while enhancing viscoelastic dissipation. Notably, *Tni* remained relatively unchanged to CoBF concentration, particularly for EDT-n samples, where values remained between 70 and 76 °C. This observation suggests that catalytic chain transfer primarily alters network connectivity and architecture rather than mesogen content or liquid-crystalline ordering. Consequently, CoBF broadens the accessible nematic temperature window by reducing *Tg* while preserving liquid-crystalline order. At equivalent catalyst loadings, EDDET-n samples consistently exhibited lower *Tg* and *E′* values than EDT-n samples because of the increased flexibility and molecular weight of the EDDET spacer.[47] These results demonstrate that both spacer structure and catalytic chain transfer provide effective routes for tuning the thermomechanical properties of methacrylate nematic elastomers. The *Tg* and *Tni* values obtained from DMA are summarized in Table 1. The *Tg* values determined by DMA are consistently higher than those obtained from DSC because the *tan δ* maximum reflects cooperative segmental relaxation on the experimental timescale, whereas DSC measures changes in heat capacity associated with the glass transition. To quantify the effect of catalytic chain transfer on network architecture, the effective crosslink density (νe) was estimated from the rubbery storage modulus using classical rubber elasticity theory:[48]

$$ve = \frac{E'}{3RT}$$

where νe is the effective crosslink density, $E'$ is the rubbery storage modulus, R is the universal gas constant, and T is the absolute temperature. The rubbery modulus was measured at 100 °C, well above the nematic-to-isotropic transition temperature. As expected, the effective crosslink density decreased systematically with increasing CoBF concentration for both network series (Table 1). For EDT-n samples, νe decreased from 135 mol $m^{-3}$ for EDT-0 to approximately 10 mol $m^{-3}$ for EDT-175. A similar trend was observed for EDDET-n, where νe decreased from 81 mol $m^{-3}$ to 53 mol $m^{-3}$. These results directly confirm that catalytic chain transfer increases the molecular weight between crosslinks and decreases the density of elastically active network strands. The reduction in crosslink density strongly influenced the viscoelastic response of the materials. Lower νe values resulted in lower *Tg* and *E′* values while simultaneously increasing t*an δ*. Because *tan δ* reflects energy dissipation within the network, these changes are particularly important for pressure-sensitive adhesion. Among all formulations, EDDET-175 exhibited the highest damping behavior, with a maximum *tan δ* of 1.33 over a broad temperature range. Notably, this formulation approaches the rheological requirements of pressure-sensitive adhesives without the need for conventional tackifiers or plasticizing additives. Together, these results demonstrate that catalytic chain transfer provides a powerful strategy for systematically tuning network architecture and thermomechanical properties in methacrylate nematic elastomers. The influence of catalytic chain transfer on the tensile behavior of the networks was further examined using uniaxial tensile testing (Figure S5). Increasing CoBF concentration generally reduced the Young's modulus and yield stress for both EDT-n and EDDET-n series, consistent with the decrease in effective crosslink density observed from DMA. The effect was particularly pronounced for EDT-n, where the initially highly crosslinked networks became substantially softer with increasing CoBF loading.

The stress–strain curves also exhibited the characteristic soft-elastic response of nematic elastomers. EDT-175 displayed a lower modulus and an extended deformation plateau compared with EDT-0 and EDT-117, reflecting the greater mobility of mesogenic segments within the less densely crosslinked network. Cyclic loading experiments (Figure S6) showed the expected recovery behavior of main-chain nematic elastomers, while EDDET-based samples containing higher CoBF concentrations exhibited improved elastic recovery owing to their lower crosslink density and increased chain flexibility.

**Table 1.** Thermomechanical properties and effective crosslink densities of methacrylate nematic elastomers as a function of catalytic chain-transfer agent concentration. The thermomechanical properties are measured using DMA

| Sample | Cobalt cat. (ppm) | *Tg* (°C) | *Tni* (°C) | *E′* measured at 100 °C (MPa) | Crosslink Density (mol/m³) |
|---|---|---|---|---|---|
| EDT-0 | 0 | 44.9 | 76.0 | 1.26 | 135.28 |
| EDT-117 | 117 | 25.9 | 70.8 | 0.09 | 9.67 |
| EDT-175 | 175 | 23.0 | 71.9 | 0.09 | 9.67 |
| EDDET-0 | 0 | 21.2 | 57.0 | 0.75 | 80.58 |
| EDDET-117 | 117 | 11.0 | 37.3 | 0.69 | 74.14 |
| EDDET-175 | 175 | 8.0 | 36.0 | 0.49 | 52.65 |

**Formulation Optimization toward Pressure-Sensitive Adhesion:**

Although catalytic chain transfer substantially reduced the crosslink density and improved the viscoelastic properties of methacrylate nematic elastomers, the resulting networks remained outside the optimal rheological window required for pressure-sensitive adhesion. To further reduce the storage modulus while maintaining liquid-crystalline order, additional monofunctional and difunctional methacrylate comonomers were incorporated into the EDDET-175 formulation, which exhibited the lowest *Tg* and highest damping behavior among the CoBF-modified networks. Figure 3a shows the chemical structures of the reactive additives used for formulation optimization. Methyl methacrylate (MMA) was selected as an isotropic monofunctional methacrylate, while CN was employed as a monofunctional liquid-crystalline methacrylate. A detailed synthesis of the CN mesogen is provided in Scheme S3 in the Supporting Information Poly(ethylene glycol) dimethacrylate (PEGDMA) was introduced as a flexible difunctional crosslinker. A constant CoBF concentration of 175 ppm was used in all formulations because this loading provided the largest reduction in crosslink density while maintaining free-standing networks. The thermomechanical properties of formulations containing 10 wt% reactive additive are shown in Figure 3b. Incorporation of the monofunctional methacrylates MMA and CN reduced the storage modulus relative to EDDET-175, consistent with a decrease in effective crosslink density arising from replacement of dimethacrylate network-forming units with monofunctional species. However, only modest changes in *Tg* were observed. In contrast, PEGDMA produced a substantial reduction in *Tg* while maintaining a similar modulus, reflecting the increased flexibility of the PEG backbone. These results indicate that monofunctional and difunctional methacrylates influence the network through different mechanisms: MMA and CN primarily reduce network connectivity through dilution of the dimethacrylate crosslinker concentration, whereas PEGDMA primarily increases chain flexibility. To simultaneously reduce both *Tg* and $E'$, formulations containing 20 wt% total additive were subsequently prepared (Figure 3c). Increasing the PEGDMA content further lowered *Tg* but produced fully isotropic materials at room temperature, eliminating the nematic order required for switchable adhesion (Figure S9). In contrast, formulations containing 10 wt% PEGDMA combined with 10 wt% MMA or CN retained liquid-crystalline behavior while exhibiting substantially improved rheological properties. Among all formulations, the network containing 80 wt% RM257–EDDET–IEM, 10 wt% PEGDMA, and 10 wt% MMA displayed the most favorable balance of properties, exhibiting a $Tg$ near 0 °C and a room-temperature storage modulus of approximately 0.3 MPa (Table S3). Consequently, this formulation most closely satisfies the Dahlquist criterion and was selected for subsequent adhesion studies.

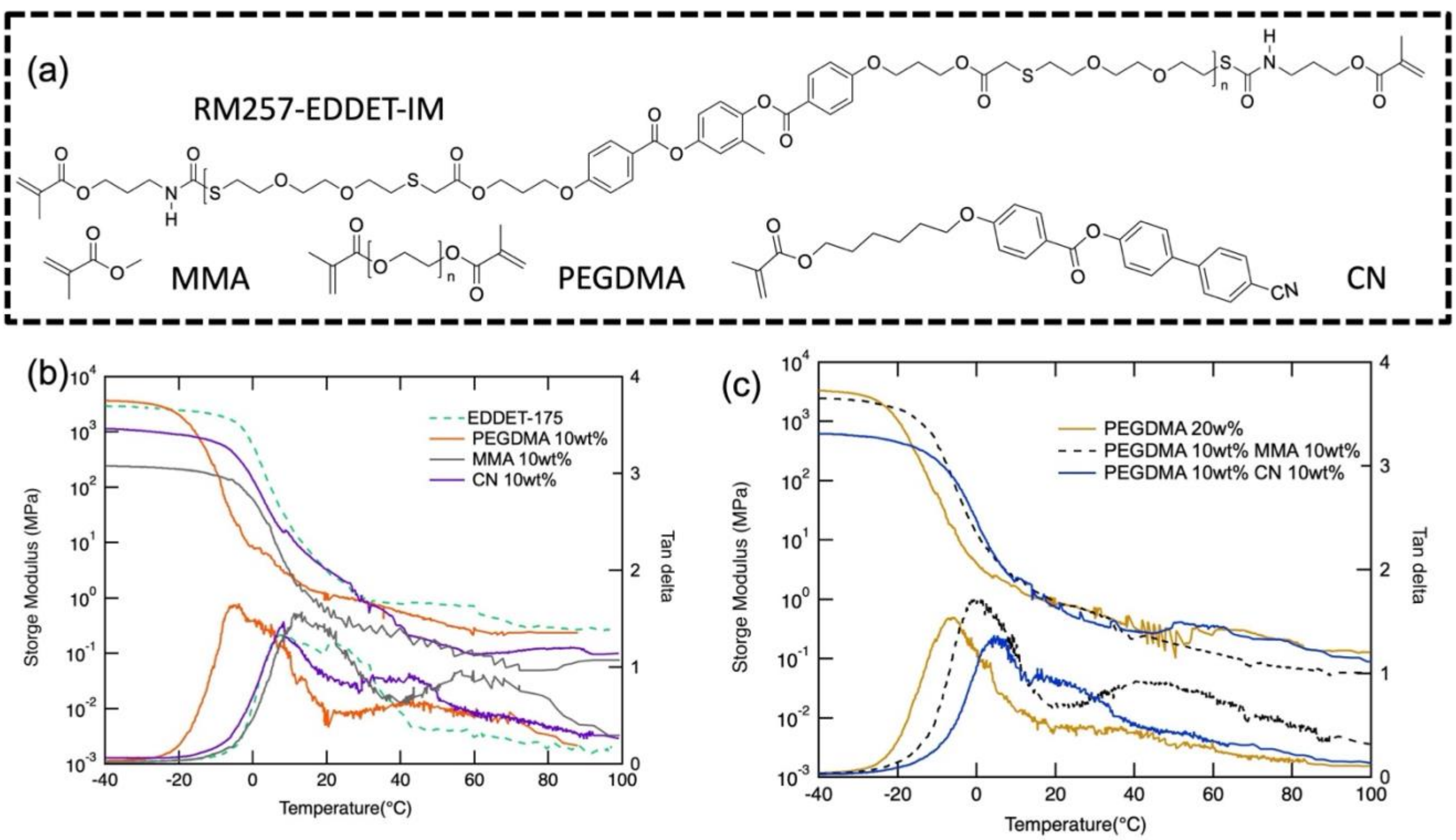

**Figure 3.** (a) Chemical structures of the reactive additives used to modify the EDDET-175 network, including the monofunctional methacrylates methyl methacrylate (MMA) and 4′-cyano-[1,1′-biphenyl]-4-yl 4-((6-(methacryloyloxy)hexyl)oxy)benzoate (CN), and the flexible difunctional crosslinker poly(ethylene glycol) dimethacrylate (PEGDMA). (b) Temperature-dependent *E′* (left axis) and *tan δ* (right axis) for formulations containing 90 wt% RM257–EDDET–IEM and 10 wt% MMA, CN, or PEGDMA. (c) Temperature-dependent *E′,* (left axis) and *tan δ* (right axis) for formulations containing 80 wt% RM257–EDDET–IEM and a total of 20 wt% reactive additive.

**Switchable Pressure-Sensitive Adhesion:**

A defining feature of switchable adhesives is the ability to combine strong adhesion under service conditions with rapid debonding when triggered by an external stimulus.[49] In nematic elastomer adhesives, this switching behavior originates from the reversible nematic-to-isotropic phase transition. In the nematic phase, the networks exhibit high viscoelastic dissipation, reflected by elevated $\tan\delta$ values, which promotes pressure-sensitive adhesion. Upon heating above $Tni$, the loss of liquid-crystalline order results in a substantial decrease in $\tan\delta$ and adhesive performance.[8,50] To evaluate the relationship between thermomechanical properties and adhesive performance, probe tack measurements were performed on formulations exhibiting low glass transition temperatures and reduced storage moduli, including EDDET-175, EDDET-175 + 10 wt% PEGDMA, EDDET-175 + 20 wt% PEGDMA, EDDET-175 + 10 wt% PEGDMA + 10 wt% CN, and EDDET-175 + 10 wt% PEGDMA + 10 wt% MMA (Figure 4a). The measured tack forces ranged from approximately 1 to 4 N and strongly depended on formulation composition. Among all materials examined, the EDDET-175 + 10 wt% PEGDMA + 10 wt% MMA formulation exhibited the highest tack force, consistent with its favorable combination of low *Tg* and low room-temperature storage modulus. In contrast, the formulation containing 20 wt% PEGDMA displayed the weakest tack performance because the material was fully isotropic at room temperature and exhibited reduced viscoelastic dissipation. Interestingly, EDDET-175 exhibited comparatively strong tack despite possessing a higher $T_g$ (8 °C) than the optimized formulations. This behavior likely arises from the enhanced contribution of nematic ordering to energy dissipation, highlighting the importance of liquid-crystalline structure in addition to conventional rheological parameters. To investigate thermally induced adhesion switching, temperature-dependent probe tack measurements were performed on EDDET-175 and EDDET-175 + 10 wt% PEGDMA + 10 wt% MMA (Figure 4b). Both materials exhibited tack forces of approximately 3.5 N at room temperature, which progressively decreased upon heating above $T_{ni}$. At 140 °C, the measured tack force approached 0.5 N, corresponding to an approximately sevenfold reduction in adhesive strength. This pronounced decrease confirms that the nematic-to-isotropic transition provides an effective mechanism for on-demand debonding. The practical implications of this switching behavior are demonstrated in Figure 4c–e and the video in the supporting information). A free-standing adhesive strip supported a 500 g load under ambient conditions in the nematic state but released the load upon heating into the isotropic phase. The clean release and absence of adhesive residue highlight the potential of these materials for reusable adhesives, reconfigurable assemblies, and debond-on-demand technologies.

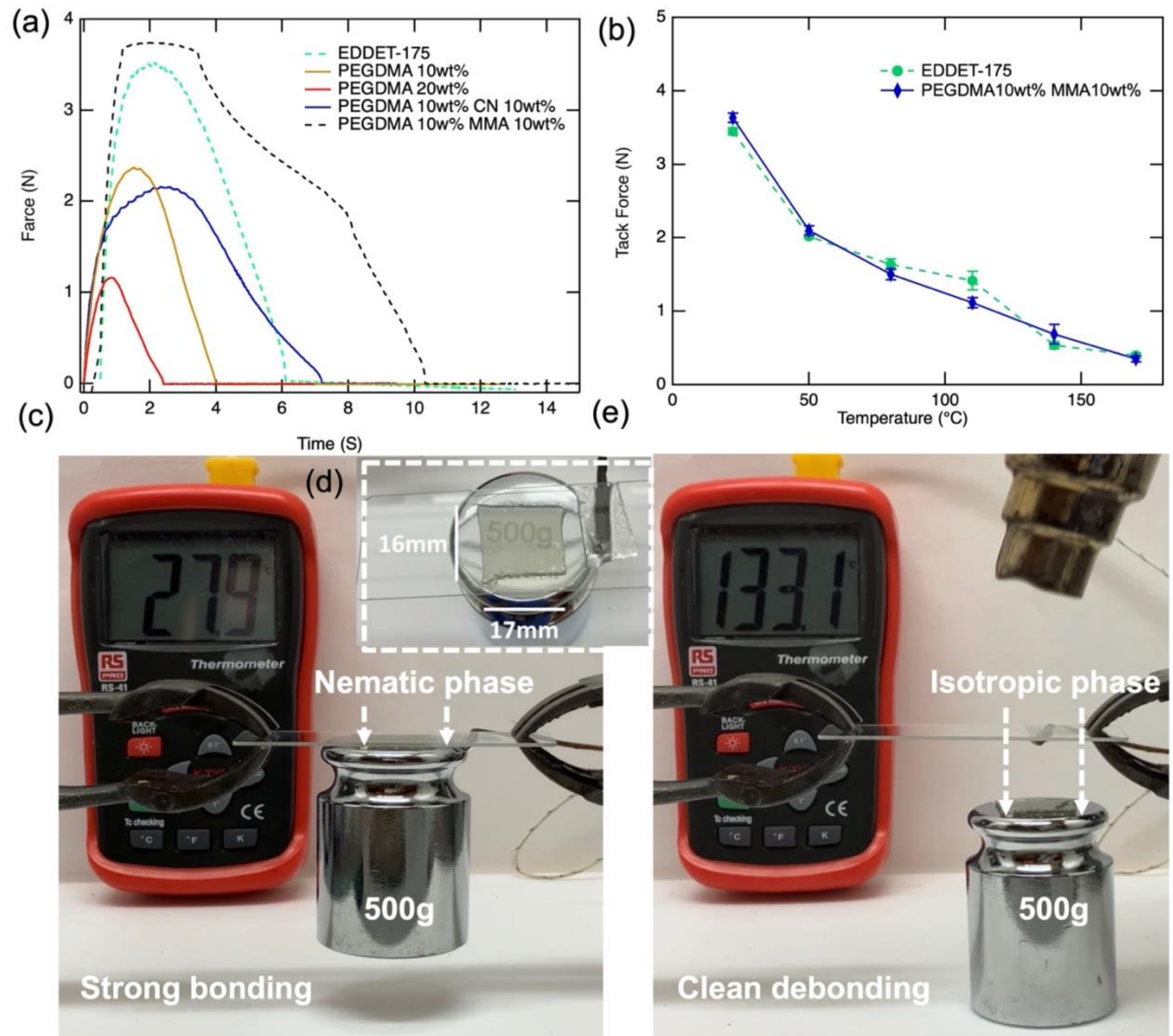

**Figure 4.** Switchable pressure-sensitive adhesion of methacrylate nematic elastomers. (a) Probe tack force measured at room temperature for EDDET-175 and modified EDDET-based formulations containing MMA, CN, and PEGDMA. (b) Temperature-dependent probe tack force for EDDET-175 and EDDET-175 + 10 wt% PEGDMA + 10 wt% MMA. (c) A free-standing adhesive strip (17 mm × 16 mm × 1 mm) bonded to a glass substrate supporting a 500 g load in the nematic state. (d) Debonding of the adhesive upon heating above the nematic-to-isotropic transition temperature. (e) Clean release of the load and adhesive from the substrate in the isotropic state without residue.

To further evaluate the practical adhesive performance of the optimized nematic elastomer formulations, temperature-dependent peel and lap-shear tests were performed (Figure 5). Unlike probe tack measurements, which primarily assess short-time interfacial adhesion, peel and lap-shear tests probe the ability of an adhesive to resist tensile and shear loading under more application-relevant conditions. The temperature dependence of adhesion was first examined using a 180° peel test (Figure 5a). At room temperature, the optimized formulation containing 10 wt% PEGDMA and 10 wt% MMA exhibited a peel strength of approximately 380 ± 19 N $m^{-1}$, whereas EDDET-175 exhibited a peel strength of 278 ± 11 N $m^{-1}$. The higher peel strength of the optimized formulation is consistent with its lower storage modulus and improved compliance, which promote close contact formation and better surface wetting and energy dissipation at the adhesive–substrate interface. Upon heating above the nematic-to-isotropic transition temperature, the peel strength of both formulations decreased dramatically and approached zero at 140 °C. This sharp reduction in adhesion reflects the substantial decrease in viscoelastic dissipation accompanying the loss of nematic order.

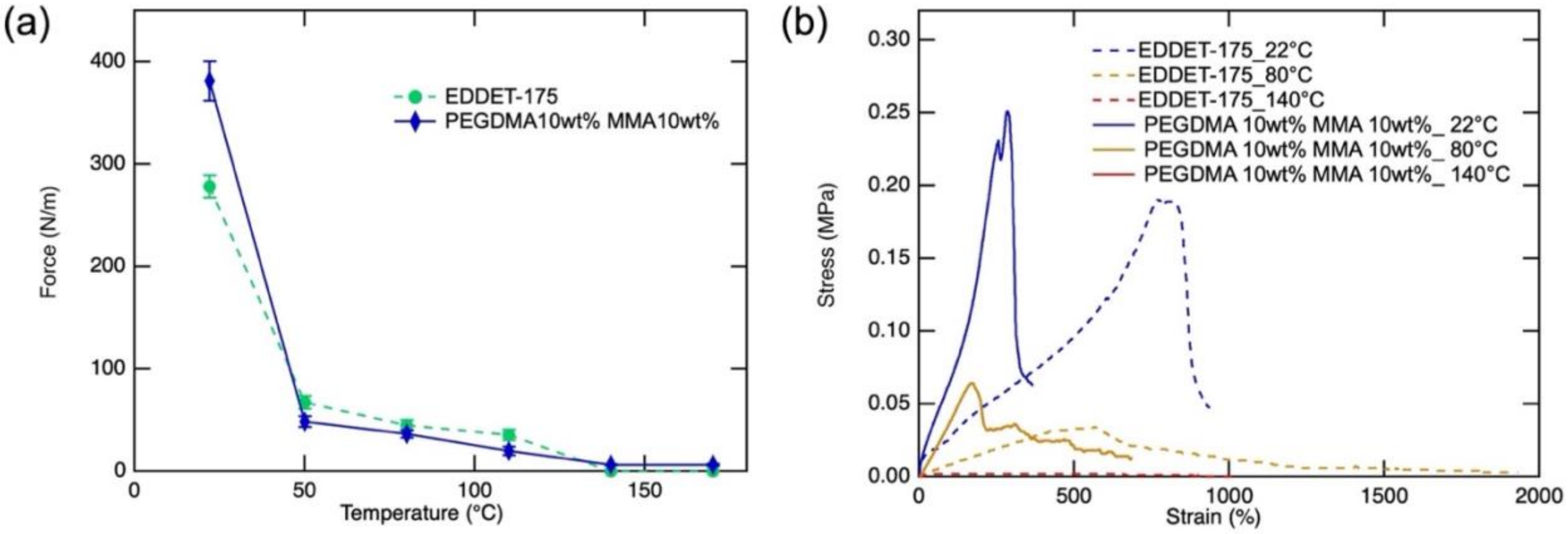

**Figure 5.** Temperature-dependent peel and lap-shear adhesion of methacrylate nematic elastomers. (a) 180° peel strength as a function of temperature for EDDET-175 and EDDET-175 + 10 wt% PEGDMA + 10 wt% MMA. Insets show reversible adhesion and thermally induced debonding of a PET-backed adhesive tape bonded to a glass substrate. (b) Lap-shear strength as a function of temperature for EDDET-175 + 10 wt% PEGDMA + 10 wt% MMA. Adhesion decreases markedly above the nematic-to-isotropic transition temperature, enabling reversible debonding on demand.

Importantly, the thermally induced debonding process was fully reversible. A PET-backed adhesive tape prepared from the optimized formulation readily supported an external load when bonded to glass at room temperature, (see video 2**,** supporting information). Heating above the nematic-to-isotropic transition resulted in rapid debonding, while simple reattachment under gentle finger pressure followed by cooling restored the original adhesive performance. This behavior demonstrates the potential of these materials as reusable pressure-sensitive adhesives capable of multiple bonding and debonding cycles. The resistance of the optimized adhesive to shear deformation was subsequently evaluated using lap-shear testing (Figure 5b). At room temperature, the formulation containing 10 wt% PEGDMA and 10 wt% MMA exhibited a lap-shear strength of approximately 0.25 MPa. Similar to the peel measurements, the shear strength decreased significantly upon heating above $T_{ni}$, falling to approximately 0.06 MPa at 80 °C and approaching zero at 140 °C. The simultaneous reduction in tack, peel strength, and lap-shear strength confirms that the adhesion-switching behavior is governed by the thermomechanical changes associated with the nematic-to-isotropic transition. Together, these results demonstrate that methacrylate nematic elastomers engineered through catalytic chain transfer can achieve strong adhesion under ambient conditions while enabling rapid, reversible, and residue-free debonding upon thermal activation. Such behavior is attractive for reusable adhesives, reconfigurable assemblies, and debond-on-demand technologies.

## CONCLUSION

In summary, we demonstrate that catalytic chain-transfer polymerization provides an effective strategy for engineering soft methacrylate nematic elastomers through systematic control of network architecture. Incorporation of ppm-level concentrations of bis(boron difluorodimethylglyoximate)cobalt(II) (CoBF) reduced the effective crosslink density, increased the molecular weight between crosslinks, and enhanced chain mobility, resulting in substantial decreases in both the glass transition temperature and storage modulus while preserving nematic order. The resulting structure–property relationships enabled precise tuning of the thermomechanical behavior of methacrylate nematic elastomers. Building upon these insights, incorporation of monofunctional methacrylates and flexible PEGDMA crosslinkers further optimized the rheological properties toward the pressure-sensitive adhesive regime. The formulation containing 80 wt% RM257–EDDET–IEM, 10 wt% PEGDMA, and 10 wt% MMA exhibited the most favorable combination of low glass transition temperature, reduced storage modulus, and high viscoelastic dissipation, approaching the Dahlquist criterion for pressure-sensitive adhesion. The optimized networks exhibited strong tack, peel, and lap-shear adhesion in the nematic state while undergoing rapid and reversible debonding upon heating above the nematic-to-isotropic transition temperature. These results establish catalytic chain transfer as a versatile platform for tailoring the thermomechanical properties of methacrylate nematic elastomers and provide a scalable route toward reusable, switchable, and debond-on-demand adhesive materials.

**ASSOCIATED CONTENT**

**Supporting Information**:

The following files are available free of charge.

Synthesis, NMR and FTIR spectra, gel fraction, DSC thermograms, DMA data, and peel-testing results (PDF).

Adhesive demonstrations comprise tack-force evaluation (Movie S1) and lap-shear testing (Movie S2).

**AUTHOR INFORMATION**

**Corresponding Author**

Mohand O Saed- Cavendish Laboratory, University of Cambridge, J.J. Thomson Avenue, Cambridge, CB3 0HE, United Kingdom

Email: mos29@cam.ac.uk

**Author Contributions**

The manuscript was written through contributions of all authors. All authors have given approval to the final version of the manuscript.

**Notes**

The authors declare that the University of Cambridge has filed a patent application related to the inventions described in this manuscript.

**ACKNOWLEDGMENT**

This work was supported by the Royal Society through University Research Fellowships (Grant Number: G113670). N.K. acknowledges support from Toray Industries. M.O. S and N.K thank Prof. Eugene Terentjev and Dr. Takuya Ohzono for their valuable discussion.